\documentstyle[preprint,aps]{revtex}
\begin{document}
\draft
\preprint{\vbox{
\hbox{CTP-TAMU-79/93}
\hbox{hep-ph/9312312}
\hbox{December 1993}
}}

\title{
Generation non-universality and flavor changing neutral currents
in the 331 model
}

\author{James T. Liu}
\address{
Center for Theoretical Physics, Department of Physics\\
Texas A\&M University, College Station, TX 77843--4242
}
\maketitle
\begin{abstract}
The 331 model, an extension of the standard electroweak theory to
$SU(3)_L\times U(1)_X$, naturally predicts three families of quarks
and leptons via the requirement of anomaly cancellation.  This is
accomplished by making one of the quark families transform differently from
the other two, thus leading to flavor changing neutral currents.  Using
experimental input on neutral meson mixing, we show that the third family
must be the one that is singled out, at least up to small family mixing.
We additionally describe a convenient way to parametrize the new mixing
matrix that plays a role in the gauge interactions of the ordinary quarks
with the new 331 quarks.
\end{abstract}

\narrowtext

%%\section{Introduction}

The 331 model is an $SU(3)_L\times U(1)_X$ extension of the standard
$SU(2)_L\times U(1)_Y$ electroweak theory \cite{pleitez,frampton}.  This
model attempts to answer the family replication question by relating the
number of families to the number of colors via the requirement of anomaly
cancellation.  While anomalies cancel for each individual family in the
Standard Model (SM), they only vanish in the 331 model when all {\it three}
families are included.
This novel method of anomaly cancellation requires that at least one family
transforms differently from the others, thus breaking generation
universality.  A result of this is that the 331 model suffers from
potentially large flavor changing neutral currents (FCNC).  Unlike generic
grand unified theories where FCNCs may be suppressed by large masses, there
is an upper bound on the scale of 331 physics on the order of a few TeV
\cite{frampton,daniel}.

The 331 model predicts several new gauge bosons beyond the SM.  These are a
new neutral gauge boson $Z'$ and a dilepton gauge boson doublet
$(Y^{++},Y^+)$.  Both the $Z'$ and new neutral scalars may have flavor
changing interactions with the quarks.  Since the leptons are generation
universal, they couple diagonally to the $Z'$ (however tree level dilepton
exchange may be lepton flavor violating \cite{lflavor}).  Thus $Z'$ FCNC is
present only in the hadronic sector of the 331 model.

Previous analyses of $Z'$ FCNC contributions to neutral meson mass
splittings have attempted to put a lower bound on the allowed $Z'$ mass
\cite{pleitez}.  However, it has since been realized that unknown mixing
parameters beyond the ordinary CKM matrix prevent one from making
quantitative statements about such a lower bound \cite{daniel,ppfcnc}.
In this paper, we show that while $Z'$ FCNC constraints
do {\it not} rule out the 331 model, the theoretical upper bound
on the $Z'$ mass may instead be used to greatly restrict the unknown mixing
parameters.  This is essentially the opposite approach from that taken
previously \cite{pleitez,daniel,ppfcnc}.  Additionally, we clarify some
of the confusion over whether the first or the third family of
quarks must be taken to transform differently.

%%\section{fermion representations}
In order to understand the origin of the $Z'$ FCNC in the 331 model, we
begin by describing the fermion representations.
While all three lepton families are treated identically, anomaly cancellation
requires that one of the three quark families transform differently from
the other two \cite{pleitez,frampton}.  In particular, cancelling the
pure $SU(3)_L$ anomaly requires that there are the same number of triplets
as anti-triplets.  Putting the three lepton families in as anti-triplets,
and taking into account the three quark colors, we find that two families of
quarks must transform as triplets and the third must transform as an
anti-triplet.

In terms of weak eigenstates, we do not need to distinguish which family
falls in the anti-triplet.  However, as we demonstrate later, it is
convenient to think of the different family as the third family.  We thus
denote the first two families as
\begin{equation}
Q_{1,2}=\pmatrix{u\cr d\cr D}_L,\quad\pmatrix{c\cr s\cr S}_L\ ,
\end{equation}
and the third family (anti-triplet) as
\begin{equation}
Q_3=\pmatrix{b\cr-t\cr T}_L\ .
\end{equation}
The sign ensures that the $SU(2)_L$ quark doublet, when embedded in $SU(3)_L$,
has the conventional form.

Using the standard normalization of non-abelian generators, the
hypercharge is embedded in $SU(3)_L\times U(1)_X$ as $Y/2=\sqrt{3}T^8+X$
where $X$ is the $U(1)_X$ charge of the representation (we use the
conventions of \cite{lflavor}).  The electric charge is then given by
$Q=T^3+Y/2$ so that $Y$ corresponds to twice the average charge of $SU(2)_L$
representations.  From these relations between $Q$, $Y$ and $X$, we find
that the electric charge of each component of $SU(3)_L$ (anti-)triplets
changes by exactly one unit and that the $X$ charge is given by the average
of the electric charges (which is just the electric charge of the middle
component).  Thus the triplets, $Q_1$ and $Q_2$ above, have $X=-1/3$ and the
third quark family, $Q_3$, has $X=2/3$.

There are then three new quarks which we denote $D$ and $S$ with electric
charge $-4/3$ and $T$ with electric charge $5/3$.  Note that all fermion
states given here are weak interaction eigenstates and must be related to
physical (mass) eigenstates by the appropriate unitary transformations.
Unlike the left-handed representations, all right handed quarks are
incorporated as $SU(3)_L$ singlets.  As a result, the ordinary right handed
quarks are generation universal and hence $Z'$ FCNC is limited to the left
handed interactions.

%%\section{The $Z'$ neutral current and FCNC}

When 331 is broken to the SM, the neutral gauge bosons $W_\mu^8$ and
$X_\mu$ mix to give the $Z_\mu'$ and hypercharge $B_\mu$ gauge bosons.
This mixing may be parametrized by a 331 mixing angle $\theta_{331}$
(generalizing the Weinberg angle) defined by \cite{lflavor}
\begin{equation}
g'={1\over\sqrt{3}}g\cos\theta_{331}={1\over\sqrt{6}}g_X\sin\theta_{331}\ ,
\end{equation}
where $g$ and $g_X$ are the $SU(3)_L$ and $U(1)_X$ coupling constants, and
the hypercharge coupling constant $g'$ is given by $\tan\theta_W=g'/g$.
In terms of $W_\mu^8$ and $X_\mu$, the hypercharge and $Z_\mu'$ gauge bosons
are given by a rotation parametrized by $\theta_{331}$
\begin{equation}
\pmatrix{B_\mu\cr Z'_\mu}
=
\pmatrix{\cos\theta_{331}&\sin\theta_{331}\cr
-\sin\theta_{331}&\cos\theta_{331}}
\pmatrix{W_\mu^8\cr X_\mu}\ .
\end{equation}

Since the $Z'$ is a combination of $W^8$ and $X$, it couples to fermions
according to
\begin{equation}
{\cal L}={g\over\sin\theta_{331}} Z'_\mu
\overline{\psi} \gamma^\mu[\cos^2\theta_{331}Y/2-T^8]\psi\ .
\end{equation}
Using $\cos\theta_{331}=\sqrt{3}\tan\theta_W$, this may be rewritten as
\begin{equation}
{\cal L}={g\over\cos\theta_W}{1\over2\sqrt{3}\sqrt{1-4\sin^2\theta_W}}
Z'_\mu J_{Z'}^\mu\ ,
\label{eq:Lzprime}
\end{equation}
where $J_{Z'}$ is given by
\begin{equation}
J_{Z'}^\mu=\overline{\psi}\gamma^\mu
[3\sin^2\theta_W Y-2\sqrt{3}\cos^2\theta_W T^8]\psi\ .
\end{equation}

Since the value of $T^8$ is different for triplets and anti-triplets, the
$Z'$ coupling to left-handed ordinary quarks is different for the third family
and thus flavor changing.  If we assume $J_{Z'}$ has a ``standard'' form for
quark triplets, then the flavor changing interaction occurs for the third
(weak-eigenstate) family and may be written as
\begin{equation}
J_{Z'({\rm FCNC})}^\mu=-2\sqrt{3}\cos^2\theta_W\overline{q}
\gamma^\mu\gamma_L [T^8({\bf3}^*)-T^8({\bf3})]q
=2\cos^2\theta_W\overline{q}\gamma^\mu\gamma_L q\ ,
\label{eq:zpfcnc}
\end{equation}
for both up- and down-type quarks ($\gamma_L={1\over2}(1-\gamma^5)$ is the
left-handed projection operator).  Other than
in the scalar sector, this is the only tree level FCNC interaction present
since when 331 is broken to the SM, all three families of ordinary quarks are
in usual $SU(2)_L$ doublets
and thus couple in the ordinary manner to the $Z$ and photon.

The dilepton currents are also sensitive to the $SU(3)_L$ structure of the
quark representations, and hence the difference in the third family.
However, with only ordinary external quarks, these dilepton effects first
show up at loop level.  Since tree level $Z'$ FCNC presumably dominates
over loop processes, a good place to study the effects of dilepton
exchange on flavor changing interactions would be in
the process $b\to s\gamma$ which cannot occur at tree level.

%%\section{$Z$--$Z'$ mixing}
After $SU(2)_L\times U(1)_Y$ breaking, the weak eigenstate $Z$ and $Z'$ may
mix, forming mass eigenstates $Z_1$ and $Z_2$.  This mixing of the neutral
gauge bosons may be parametrized by a mixing angle $\phi$ so that
\begin{equation}
\pmatrix{Z_1\cr Z_2}
=\pmatrix{\cos\phi&-\sin\phi\cr
\sin\phi&\cos\phi}
\pmatrix{Z\cr Z'}\ .
\label{eq:zzmix}
\end{equation}
A fit to precision electroweak observables gives a limit on the mixing
angle of $-0.0006<\phi<0.0042$ and a lower bound on the mass of the heavy
$Z_2$ of $M_{Z_2}>490$GeV (both at 90\% C.L.)
\cite{zzmix}.  While this mass limit is not as strong as the indirect limit
$M_{Z_2}>1.4$TeV given by the dilepton mass bound and the $M_{Z_2}$--$M_Y$
relation of the minimal Higgs sector \cite{phenom}, it is insensitive to
the choice of Higgs representations and provides an independent
lower bound on $M_{Z_2}$.

Due to the mixing, the mass eigenstate $Z_1$ now picks up flavor changing
couplings proportional to $\sin\phi$.  In particular, using
(\ref{eq:Lzprime}) and (\ref{eq:zzmix}), we may write \cite{daniel}
\begin{equation}
{\cal L}_{\rm FCNC}={g\over\cos\theta_W}
{1\over2\sqrt{3}\sqrt{1-4\sin^2\theta_W}}
(-\sin\phi Z_{1\mu}+\cos\phi Z_{2\mu}) J_{Z'({\rm FCNC})}^\mu\ .
\label{eq:z12fcnc}
\end{equation}
For sufficiently large mixing, the flavor changing $Z_1$ decays may be
observable.  However, because $Z$--$Z'$ mixing is constrained to be very
small, evidence of 331 FCNC can only be probed indirectly at present via the
$Z_2$ couplings.

%%\section{quark masses and mixing}
In order to examine the flavor changing $Z'$ interaction given in
(\ref{eq:zpfcnc}), we need to relate weak and mass eigenstate quarks.
Symmetry breaking and mass generation in the minimal 331 model is
accomplished by four Higgs multiplets --- the three triplets
\begin{equation}
\Phi=\pmatrix{\Phi_Y\cr\varphi^0}\qquad
\phi=\pmatrix{\Phi_1\cr\Delta^-}\qquad
\phi'=\pmatrix{\widetilde{\Phi}_2\cr\rho^{--}}\ ,
\end{equation}
with $X$ charges 1, 0, and $-1$ respectively and a sextet $H$ with $X=0$
\cite{frampton,daniel,lflavor,foot}.  We have written the triplets in terms
of $SU(2)_L$ component fields:  $\Phi_Y=(\Phi_Y^{++},\Phi_Y^+)^T$, the
Goldstone boson doublet corresponding to the massive dileptons and
$\Phi_i=(\phi_i^+,\phi_i^0)^T$, which are SM Higgs doublets where
$\widetilde{\Phi}_i=i\tau^2\Phi_i^*$.  A third SM doublet arises from the
sextet $H$, but plays no role in generating quark masses.

The vacuum expectation value $\langle\Phi\rangle$ breaks 331 and gives masses
to the new quarks
$D$, $S$, and $T$.  The remaining scalars implement $SU(2)_L\times U(1)_Y$
breaking and gives masses to the remaining fermions.
In particular, the most general gauge invariant Yukawa couplings of the
above scalars to the quarks may be written
\begin{eqnarray}
-{\cal L}=&&\overline{Q_{Li}'}h_d^{ik}d_{Rk}'\phi+\overline{Q_{L3}'}
h_d^{3k}d_{Rk}'\phi'^*\nonumber\\
&+&\overline{Q_{Li}'}h_u^{ik}u_{Rk}'\phi'-\overline{Q_{L3}'}h_u^{3k}
u_{Rk}'\phi^*\nonumber\\
&+&\overline{Q_{Li}'}h_D^{ij}D_{Rj}'\Phi+\overline{Q_{L3}'}h_T
T_R^{\vphantom{\prime}}\Phi^*
+{\rm h.c.}\ ,
\end{eqnarray}
where $i,j=1,2$ runs through the first two families only and $k=1,2,3$.  As
usual, the primes denote weak eigenstates.  Since $T$ is the only charge
$5/3$ quark, it is a simultaneous gauge and mass eigenstate.

When 331 is reduced to the SM, the Yukawa interactions may be written in
terms of ordinary left-handed quark doublets $q_{Li}=(u_i,d_i)_L^T$ and
singlets.  We separate ${\cal L}$ into two pieces, ${\cal L}_0$ which
contains only lepton number $L=0$ scalars and ${\cal L}_2$ which has $|L|=2$
scalars that change ordinary and new quarks into each other.  We find
\begin{eqnarray}
-{\cal L}_0=&&\overline{q_{Li}'}h_d^{ik}d_{Rk}'\Phi_1+\overline{q_{L3}'}
h_d^{3k}d_{Rk}'\Phi_2\nonumber\\
&+&\overline{q_{Li}'}h_u^{ik}u_{Rk}'\widetilde{\Phi}_2
+\overline{q_{L3}'}h_u^{3k}u_{Rk}'\widetilde{\Phi}_1\nonumber\\
&+&\overline{D_{Li}'}h_D^{ij}D_{Rj}'\varphi^0+
\overline{T_L^{\vphantom{\prime}}} h_TT_R^{\vphantom{\prime}}
\varphi^{0*}+{\rm h.c.}\nonumber\\[4pt]
-{\cal L}_2=&&\overline{q_{Li}'}h_D^{ij}D_{Rj}'\Phi_Y-\overline{q_{L3}'}h_T
T_R^{\vphantom{\prime}}\widetilde{\Phi}_Y\nonumber\\
&+&\overline{D_{Li}'}h_d^{ik}d_{Rk}'\Delta^-+\overline{D_{Li}'}h_u^{ik}
u_{Rk}'\rho^{--}\nonumber\\
&+&\overline{T_L^{\vphantom{\prime}}}h_d^{3k}d_{Rk}'\rho^{++}
-\overline{T_L^{\vphantom{\prime}}}h_u^{3k}u_{Rk}'
\phi'^++{\rm h.c.}\ .
\label{eq:qyuk2}
\end{eqnarray}
Because the third family of quarks is treated differently, it has different
couplings to scalars as well as the $Z'$.  Thus natural flavor conservation
\cite{glashow} is necessarily violated in the 331 model, leading to
potentially large flavor changing neutral Higgs (FCNH) processes in addition
to $Z'$ FCNC.

If it were not for the third
family, the ordinary quarks in ${\cal L}_0$ would have a normal two Higgs
doublet coupling with separate Higgs couplings to up- and down-type quarks
(usually referred to as model II).  The third family, however, has a
``flipped'' coupling, with $\Phi_1$ and $\Phi_2$ exchanging roles.
Including the $SU(3)_L$ sextet Higgs which give masses to the leptons,
we end up with a three-Higgs doublet model, albeit with unusual Yukawa
couplings dictated by the underlying $SU(3)_L$ theory.

Since the $Z'$ couples differently to the third weak-eigenstate family, $Z'$
FCNC occurs through of a mismatch between weak and mass eigenstates.  Since
there are more states than in the SM, this mixing is described by more than
just the CKM matrix.  The charge $2/3$, $-1/3$ and $-4/3$ mass matrices are
diagonalized by three independent bi-unitary transformations which we denote
by the $3\times3$ unitary matrices $U_{L,R}$ and $V_{L,R}$ and the
$2\times2$ unitary matrices $W_{L,R}$ respectively.  In the standard
fashion, the ordinary CKM matrix is given by $V_{CKM}=U_L^\dagger
V_L^{\vphantom{\dagger}}$.
The new mixing shows up in both dilepton currents and the FCNC part of the
$Z'$ interaction.

Because the first two families are generation symmetric, we may make a
convenient choice of letting $D$ and $S$ be simultaneous gauge and mass
eigenstates.  This replaces the standard choice of using up-type quarks in
this fashion which is no longer possible in this case.  As a result, the
charged currents in the quark sector and the $Z'$ FCNC interaction may be
written
\begin{eqnarray}
J_{W^+}^\mu&=&\overline{u}\gamma^\mu\gamma_LV_{CKM}d\nonumber\\
J_{Y^+}^\mu&=&\overline{d}\gamma^\mu\gamma_LV_L^\dagger
\pmatrix{1&0\cr0&1\cr0&0}D+\overline{T}\gamma^\mu\gamma_L\pmatrix{0&0&1}
U_Lu\nonumber\\
J_{Y^{++}}^\mu&=&\overline{u}\gamma^\mu\gamma_LU_L^\dagger
\pmatrix{1&0\cr0&1\cr0&0}D-\overline{T}\gamma^\mu\gamma_L\pmatrix{0&0&1}
V_Ld\nonumber\\
J_{Z'(FCNC)}^\mu&=&2\cos^2\theta_W[\overline{u}\gamma^\mu\gamma_L
U_L^\dagger\pmatrix{0&&\cr&0&\cr&&1}U_Lu+\overline{d}\gamma^\mu\gamma_L
V_L^\dagger\pmatrix{0&&\cr&0&\cr&&1}V_Ld]\ ,
\label{eq:qcc}
\end{eqnarray}
in a matrix notation where $D=(D,S)^T$.  If we had not initially picked $D$ to
be generation diagonal, we could simply have absorbed the unitary matrix $W_L$
into a redefinition of $U_L$ and $V_L$.

Unlike the SM where only $V_{CKM}$ is physical, there is additional freedom
in the mixing present above \cite{ppfcnc}.  Although we have introduced
three matrices in (\ref{eq:qcc}), they are not independent but are
related by $V_{CKM}=U_L^\dagger V_L^{\vphantom{\dagger}}$.  Since flavor
changing interactions involving down-type quarks have been studied the
most extensively, we find it convenient to specify the two unitary matrices
$V_{CKM}$ and $V_L$.  As usual, $V_{CKM}$ contains three angles and one
complex phase.  $V_L$ is specified by three angles and three phases since we
may remove three phases from the general unitary matrix by appropriately
transforming the three new quarks.

In the absence of CP violating phases, the three angles of $V_L$ have a
simple interpretation.  We may use a CKM like parametrization
\begin{equation}
V_L=\pmatrix{v_{1d}&v_{1s}&v_{1b}\cr v_{2d}&v_{2s}&v_{2b}\cr
v_{3d}&v_{3s}&v_{3b}}
=\pmatrix{c_{12}c_{13}&-s_{12}c_{23}-c_{12}s_{23}s_{13}
&s_{12}s_{23}-c_{12}c_{23}s_{13}\cr
s_{12}c_{13}&c_{12}c_{23}-s_{12}s_{23}s_{13}
&-c_{12}s_{23}-s_{12}c_{23}s_{13}\cr
s_{13}&s_{23}c_{13}&c_{23}c_{13}}\ ,
\label{eq:param}
\end{equation}
where $s_{ij}=\sin\theta_{ij}$ and $c_{ij}=\cos\theta_{ij}$.  Since the
third row corresponds to the anti-triplet weak eigenstate, $\theta_{13}$
and $\theta_{23}$ specify which down-type quark is in the anti-triplet and,
orthogonal to that, $\theta_{12}$ specifies the mixing between the first
two triplets ({\it i.e.}~$D$ and $S$).

%%\section{neutral meson mixing and restrictions on $V_L$}
Previous examinations of the $Z'$ in the 331 model have concentrated on
putting lower bounds on $M_{Z_2}$ \cite{pleitez,ppfcnc} to prevent
excessive tree-level FCNC.  The drawback to this approach is that the new
mixing specified by $V_L$ is in principle unknown and has to be estimated.
Here, we instead use the {\it upper} limit $M_{Z_2}<2.2$GeV \cite{phenom}
to place restrictions on $V_L$.

The strongest constraints on tree level $Z'$ FCNC come from neutral meson
mixing.  For the neutral kaon system, the tree level $\Delta S=2$
interaction is given from (\ref{eq:zpfcnc}) and (\ref{eq:z12fcnc}) by
\begin{equation}
{\cal H}_{\rm eff}^{\Delta S=2}
={g^2 c^2\over12(1-4s^2)}(v_{3s}^*v_{3d}^{\vphantom{*}})^2
\left({\cos^2\phi\over M_{Z_2}^2}+{\sin^2\phi\over M_{Z_1}^2}\right)
[\overline{s}\gamma^\mu\gamma_Ld][\overline{s}\gamma_\mu\gamma_Ld]
\end{equation}
where $s=\sin\theta_W$ and $c=\cos\theta_W$.  In addition to the SM box
diagram and possible long distance effects, this contributes a term
\begin{equation}
\Delta m_K={2\sqrt{2}\over9}G_F{c^4\over (1-4s^2)}
|v_{3s}^*v_{3d}^{\vphantom{*}}|^2
\left(\eta_{Z_2}\cos^2\phi{M_{Z_1}^2\over M_{Z_2}^2}+\eta_{Z_1}\sin^2\phi
\right) B_Kf_K^2m_K\ ,
\end{equation}
to the $K^0$--$\overline{K^0}$ mass difference \cite{fn:cpvio}.
We have included the leading order QCD corrections through the parameters
$\eta_{Z_2}\approx0.55$ and $\eta_{Z_1}\approx0.61$ \cite{gilman}.
$B_K$ and $f_K$ are the bag parameter and decay constant of the kaon.
Similar equations hold for $D^0$--$\overline{D^0}$ and
$B^0$--$\overline{B^0}$ mixing.  Because the $Z$--$Z'$ mixing angle $\phi$ is
very small, the first term in the parentheses dominates and $\sin^2\phi$
may safely be neglected.

The present limits on neutral meson mixing are given by \cite{pdb}
\begin{equation}
\begin{array}{lrl}
K^0\hbox{--}\overline{K^0}&\qquad\Delta m =&(3.522\pm0.016)\times
10^{-12}\rm MeV\\
D^0\hbox{--}\overline{D^0}&<&1.3\times10^{-10}\rm MeV\\
B^0\hbox{--}\overline{B^0}&=&(3.6\pm0.7)\times10^{-10}\rm MeV\ .
\end{array}
\end{equation}
Although there is considerable uncertainty in the heavy meson decay
constants, this has little effect on the results.  We use
\begin{eqnarray}
\sqrt{B_K}f_K&=&135\pm19\rm MeV\nonumber\\
\sqrt{B_D}f_D&=&187\pm38\rm MeV\nonumber\\
\sqrt{B_B}f_B&=&208\pm38\rm MeV\ .
\end{eqnarray}
The kaon quantity comes from $f_K=161$MeV and $B_K=0.7\pm 0.2$.  The
heavy meson decay constants are taken from a lattice calculation,
Ref.~\cite{lattice}, where all reported errors are added in
quadrature and $B_D=B_B=1$.

Because there are various sources that may contribute to the mass
difference, $\Delta m$, it is impossible to disentangle the tree level
$Z'$ contribution from other effects.  However, barring any unexpected
cancellations, it is reasonable to expect that $Z'$ exchange contributes a
$\Delta m$ no larger than the observed values.
Using the upper limit, $M_{Z_2}<2.2$TeV, we find, from the $K^0$, $D^0$ and
$B^0$ system, respectively
\begin{eqnarray}
|v_{3s}^*v_{3d}^{\vphantom{*}}|&<&\phantom{0}5.0\times10^{-3}\nonumber\\
|u_{3c}^*u_{3u}^{\vphantom{*}}|&<&10.8\times10^{-3}\nonumber\\
|v_{3b}^*v_{3d}^{\vphantom{*}}|&<&\phantom{0}8.7\times10^{-3}\ ,
\label{eq:fcnclim}
\end{eqnarray}
(at 90\%C.L.).  $u_{3i}$ are components of the third row of $U_L$, the
rotation matrix in the up-quark sector, and are given by
$u_{3i}=v_{3j}V^*_{CKM\,ij}$.

It should now be apparent why we have chosen to parametrize the new mixing
by $V_L$.  In this case, we make it easy to describe FCNC in the more
interesting $K^0$ and $B^0$ systems at the expense of the $D^0$.  In the
parametrization of $V_L$ given by Eq.~(\ref{eq:param}), $v_{3i}$ is
determined (in magnitude) by two angles, $\theta_{13}$ and $\theta_{23}$.
Since $|v_{3s}^*v_{3d}^{\vphantom{*}}|={1\over2}|\sin\theta_{23}
\sin2\theta_{13}|$ and $|v_{3b}^*v_{3d}^{\vphantom{*}}|={1\over2}
|\cos\theta_{23}\sin2\theta_{13}|$, we immediately
determine from (\ref{eq:fcnclim}) that $|\sin2\theta_{13}|<0.020$, giving
$|\theta_{13}+n\pi/2|<0.010$.  This allows two types of solutions, either
$|v_{3d}|\approx0$ (the second or third family is the anti-triplet)
or $|v_{3d}|\approx1$ (the first family is the anti-triplet).

In order to restrict these cases further, we must relate $u_{3i}$ to $v_{3i}$
and take the limit on $D^0$ mixing into account.  This requires knowledge of
$V_{CKM}$ and possible new CP violating phases as well.  We find that in
order to satisfy all three conditions of (\ref{eq:fcnclim}) simultaneously,
only $|v_{3d}|\approx0$ is allowed.  Restricted to the first quadrant, the
limits on $\theta_{ij}$ are
\begin{equation}
\theta_{13}<0.010,\qquad \theta_{23}<0.26\ ,
\label{eq:zpcons}
\end{equation}
which means $|v_{3b}|\approx1$ and hence that the third family must be the
anti-triplet (up to small mixing).

There has been some confusion over the issue of whether the first
family or the third family must be treated differently in order to
sufficiently suppress the $Z'$ FCNC \cite{pleitez,daniel,ppfcnc}.
Obviously, in terms of weak eigenstates, it makes no difference which family
is assigned to the anti-triplet.  In terms of mass eigenstates, the
anti-triplet has been unitarily transformed into some combination of all
three families.  However, physically, the almost-diagonal CKM matrix tells
us that it makes sense to group mass eigenstates into families.  It is in
this manner that we may say the third family must be the one that is
different.  The reason this choice is forced on us is because the Cabibbo
angle, $\sin\theta_C\approx0.22$, is the largest off-diagonal element of
$V_{CKM}$, and hence the $\Delta S=2$ and $\Delta C=2$ FCNC limits cannot
be simultaneously satisfied unless the anti-triplet is in the third family.

When $B_s^0$ mixing is measured, it will put further stronger restrictions
on $\theta_{23}$.  In the SM, $\Delta m_{B_d}/\Delta m_{B_s}\sim
|V_{CKM\,td}/V_{CKM\,ts}|^2$ so $B_s^0$ mixing is expected to be large.
Although this box diagram contribution is still present in the 331 case,
if we assume that the tree level process dominates, we find instead
$\Delta m_{B_d}/\Delta m_{B_s}\sim|v_{3d}/v_{3s}|^2=|\tan\theta_{13}/
\sin\theta_{23}|^2$.  Depending on the new mixing angles, the $Z'$
contribution to $B_s^0$ mixing may be large or small.
Even if this mixing turns out to be unexpectedly small, it will not rule out
the 331 model.  Because of the additional freedom present in $V_L$, there is
a possibility that tree level $Z'$ exchange has the opposite phase as the
SM box diagram, and hence would suppress the large SM contribution to
$\Delta m_{B_s}$.  This intriguing possibility of small $B_s^0$ mixing
would present clear evidence of physics beyond the SM, including possible
support for the 331 model.

Tree level $Z'$ exchange also contributes to $\Delta S=1$ FCNC processes
such as $K\to\pi\nu\overline{\nu}$.  We find
\begin{equation}
{{\rm BR}(K^+\to\pi^+\nu\overline{\nu})\over{\rm BR}(K^+\to\pi^0e^+
\nu_e)}={c^4\over3}{|v_{3s}^*v_{3d}^{\vphantom{*}}|^2\over|V_{CKM\, us}|^2}
\left( \cos^2\phi{M_{Z_1}^2\over M_{Z_2}^2}+ \sin^2\phi \right)^2\ .
\end{equation}
Since ${\rm BR}(K^+\to\pi^+\nu\overline{\nu})<1.7\times10^{-8}$
\cite{atiya}, we use the
upper bound on $M_{Z_2}$ to find $|v_{3s}^*v_{3d}^{\vphantom{*}}|<0.18$
which is a weaker limit than that from $K^0$--$\overline{K^0}$ mixing,
Eq.~(\ref{eq:fcnclim}).  Similar considerations hold for the rare decay
$K^0_L\to\mu^+\mu^-$.  However, it is theoretically harder to treat because
of long-distance contributions.  The reason such semi-leptonic decays do
not give strong mixing constraints is that the $Z'$ is only weakly coupled
to the leptons.

While the above processes occur at tree level via $Z'$ exchange, the rare
decay $b\to s\gamma$ must still proceed at one-loop.  In the 331 model, in
addition to the SM $W$ penguin, this may occur via $Z'$ and $Y$ penguins.
Although the SM contribution is GIM suppressed, this is no longer the case
for both 331 contributions.  One might worry that this would lead to too
large a rate for $b\to s\gamma$.  However, the non-GIM suppressed
contributions are proportional to new mixing given by
$v_{3b}^*v_{3s}^{\vphantom{*}}$, which
may be sufficiently small to prevent conflict with experiment \cite{cleo}.
This is currently under investigation \cite{afl}.

%%\section{conclusion}
In conclusion, FCNC occurs at tree level in the 331 model because of the
$Z'$, which couples differently to triplets and anti-triplets.  In order
to describe the flavor changing $Z'$ interaction, we need to understand
family mixing in the quark sector, which is complicated by the presence
of the new quarks.  In addition to the ordinary CKM matrix, three more
angles and three new phases are required to describe the mixing between
ordinary and new quarks.  Although we have not focused on the three new
CP violating phases, they may lead to striking predictions beyond the SM
and deserve further investigation.

We find that the only way to satisfy the experimental constraints on FCNC
is to make the third family transform differently from the other two (up
to small mixing).  The reason behind singling out the third family is that
it has the smallest couplings to the other two families --- the Cabibbo angle
mixing is sufficiently large that it forces the first two families to be
treated identically.  Because of the almost diagonal family structure, it
makes physical sense to group either weak or mass eigenstate quarks into
corresponding families.  This is why it is convenient to think of the third
family as unique, even in terms of weak eigenstates \cite{frampton,daniel},
although technically it makes no difference.

Going back to the quark Yukawa couplings, (\ref{eq:qyuk2}), we note
that since the Higgs couplings to the third family are different, FCNH
will occur in the scalar sector.  However, the $Z'$ FCNC constraint,
(\ref{eq:zpcons}), will simultaneously suppress FCNH by restricting the
third family to be almost diagonal.  Thus the SM Yukawa interactions are
similar to that of the two-Higgs doublet model II with the exception that
$t$ and $b$ get their masses from the opposite Higgs doublet as for the first
two families.

Because of the unique feature that there is an upper bound on the unification
scale, the 331 model is highly predictive.  It is remarkable that in this
model, there is just enough freedom to eliminate large FCNC, and the result
of this is to constrain the third family to be the one that is different.
In turn, this may give us some indication of why the top quark is so heavy
and may present a new approach to the question of fermion mass generation.

%%\section*{Acknowledgements}
\bigskip
I would like to thank D.~Ng for helpful comments and input on some of this
work, and J.~Agrawal, P.~Frampton and P.~Krastev for enlightning discussions.
This work was supported in part by the U.S.~Department of Energy under Grant
No.~DE-FG05-85ER-40219 and the National Science Foundation under Grant
No.~PHY-916593.


\begin{references}
\bibitem{fn:cpvio} For this discussion we ignore CP violating effects,
but note that ${\cal H}_{\rm eff}^{\Delta S=2}$ may also give rise to a
large $\epsilon$ parameter, depending on unknown phases.
\end{references}
\end{document}